\documentclass[%
 reprint,
 amsmath,amssymb,
 aps,
prl,
]{revtex4-1}

\usepackage{graphicx}
\usepackage{dcolumn}
\usepackage{bm}

\newcommand{\qcol}{0.24\linewidth}

\begin{document}


\title{Profile approach for recognition of three-dimensional magnetic structures}
\author{I. A. Iakovlev, O. M. Sotnikov, V. V. Mazurenko}
%
\affiliation{
Theoretical Physics and Applied Mathematics Department, Ural Federal University, Mira Street 19, Ekaterinburg 620002, Russia
}
\date{\today}



\begin{abstract}
We propose an approach for low-dimensional visualisation and classification of complex topological magnetic structures formed in magnetic materials. Within the approach one converts a three-dimensional magnetic configuration to a vector containing the only components of the spins that are parallel to the $z$ axis. The next crucial step is to sort the vector elements in ascending or descending order.  Having visualized profiles of the sorted spin vectors one can distinguish configurations belonging to different phases even with the same total magnetization. For instance, spin spiral and paramagnetic states with zero total magnetic moment can be easily identified. Being combined with a simplest neural network our profile approach provides a very accurate phase classification for three-dimensional magnets characterized by complex multispiral states even in the critical areas close to phases transitions. By the example of the skyrmionic configurations we show that profile approach can be used to separate the states belonging to the same phase.     
  
\end{abstract}

\pacs{Valid PACS appear here}
\maketitle


{\it \label{sec:level1}Introduction.}
Amazing progress in computer recognition and classification of two-dimensional images related to recent development of machine learning techniques \cite{image1,image2,image3, image4} and growth of the computer power has revolutionized the field of computer vision technologies \cite{image5, image6}. It also facilitates the solution of the new wave of challenges, such as recognition of a three-dimensional (3D) object, the problem that arises in different fields of research from modelling and decoding of the human brain \cite{Gallant} to self-driving cars \cite{self-driving}. This challenge is related to a number of fundamental problems in computer vision: representation of a 3D object, identification of the object from its image, estimation of its position and orientation, and registration of multiple views of the object for automatic model construction. There are different approaches to present 3D shapes, such as voxel grid \cite{Wu} and view-based descriptors \cite{Hang}. The situation becomes more complicated when 3D objects form a system.

In physics the image recognition problem can be reformulated as a phase recognition one. For instance, we need to classify a set of experimental or theoretical magnetic structures with respect to different ordered or disordered phases.  From the experimental side the techniques for observation of the three-dimensional spin configurations are  actively developed \cite{holography, milde} and the main focus in these studies is on the systems revealing the topologically-protected skyrmion excitations. The main challenges to be addressed before the skyrmion can be used in actual devices are the fabrication of thin films containing skyrmions at room temperature and clarification of their 3D magnetic structure \cite{holography}. 

\begin{figure}[t] 
\center 
\includegraphics[width=\columnwidth ,clip]{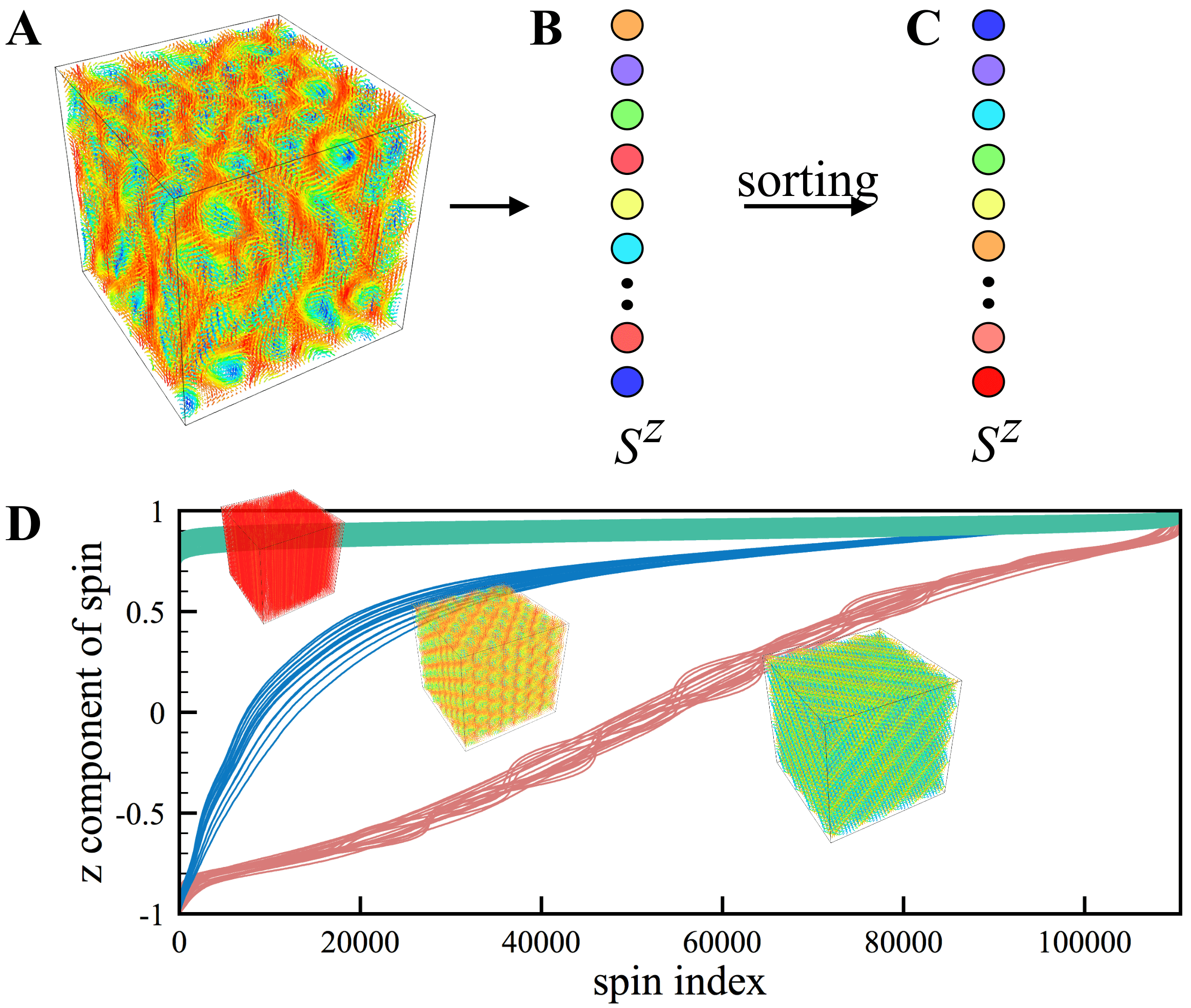}
\caption{Idea of the profile transformation. Complete magnetic structure (A) of a 3D system is reduced to unsorted vector (B) containing $S^z$ components of spins. (C) Sorted spin vector, red and blue circles correspond to the spins with $z$ projections of 1 and -1, respectively.  (D) Visualization of the magnetization profiles of the Monte Carlo-sampled configurations obtained for spin spiral (red lines), skyrmion (blue lines) and ferromagnetic (green lines) phases. Different phases can be identified by their magnetization profile.}
\label{profile1}
\end{figure}

From the theory side there is an arsenal of methods of supervised \cite{Melko1,Melko2} and unsupervised \cite{Evert,Maaten} learning that can be potentially used for classification of the three-dimensional spin structures and identification of hidden patterns in unstructured spin configurations data. However, it was not done up to date. The main complexity is related to high dimensionality of the data sets for 3D magnets. For instance, for realistic simulations of the skyrmion systems various magnetization meshes from 100$\times$100$\times$3 (30000 spins in total) \cite{FeCoSi} to 256$\times$256$\times$280 (18350080 spins in total) \cite{3dskyrmion1} were used. Here each spin has three components, which significantly increases the dimension of the problem. At the same time, advanced techniques of unsupervised learning and visualisation of high-dimensional data such as t-distributed stochastic neighbor embedding (t-SNE) \cite{Maaten} give reliable results for data of a few thousand dimensions.  

Here we report on the approach for classification of the magnetic structures formed in the three-dimensional ferromagnets with Dzyaloshinskii-Moriya interaction. We show that in the case of the magnetic systems a complex dimensional reduction that is a standard procedure of the unsupervised techniques \cite{Hotelling, Maaten} can be partially performed manually, by simply neglecting the in-plane components of the magnetization. The rest, out-of-plane components of spins is represented as one-dimensional vector (Fig.\ref{profile1} B) that is to be sorted (Fig.\ref{profile1} C) and visualized (Fig.\ref{profile1} D). These are the main steps of our profile approach to distinguish different magnetic phases and patterns in three-dimensional magnets. Fig.\ref{profile1} D gives an example of the profile transformation of the magnetic structures belonging to the skyrmion, ferromagnetic and spin spiral phases. One can immediately realize that magnetization profile has a specific shape and features for each phase, which facilitates the recognition. For instance, ferromagnetic systems are characterized by constant profile with maximal magnetization. In the case of the spin spirals the $z$ components of spins reveal nearly uniform distribution between -1 and 1. In combination with feed-forward network the profile transformation provides a very accurate description of the transitional areas between different phases. We also demonstrate the capacities and scalability of the profile approach by the example of the recognition of different skyrmionic species in 3D magnets.  

{\it \label{sec:level1} Model and Methods.}
The approach we propose is of general nature and allows one to perform the classification of magnetic structures obtained from experiments \cite{holography, milde,donnelly,Tanigaki} and simulations \cite{FeCoSi,3dskyrmion1,3dskyrmion2,3dskyrmion3}.
In this work we concentrate on Monte Carlo simulations \cite{supplementary} of the following Hamiltonian 
\begin{equation}\label{Ham}
\begin{split}
H=- \sum_{i<j}J_{ij}{\bf S}_i{\bf S}_j-\sum_{i<j}{\bf D}_{ij}[{\bf S}_i\times{\bf S}_{j}]-  \sum_i B S_i^z
\end{split}
\end{equation}
on the 48$\times$48$\times$48 cubic lattice. Here $J_{ij}$ and ${\bf D}_{ij}$ are the isotropic exchange interaction and Dzyaloshinskii-Moriya vector, respectively. ${\bf S}_{i}$ is a unit vector along the direction of the $i$th spin and $B$ denotes the $z$-oriented magnetic field. We take into account the only interactions between nearest neighbours. In our simulations the isotropic exchange interaction, $J$ = 1 is positive, which corresponds to the ferromagnetic case. All the parameters in this work are defined in units of $J$. DMI is parallel to the corresponding inter-site radius vector. Such a Hamiltonian was previously used in Ref.\onlinecite{FeCoSi} to simulate the magnetic phase diagram of the Fe$_{0.5}$Co$_{0.5}$Si thin films demonstrating topologically-protected magnetic states, skyrmions at finite temperatures and magnetic fields.

\begin{figure}[t] 
\center 
\includegraphics[width=\columnwidth ,clip]{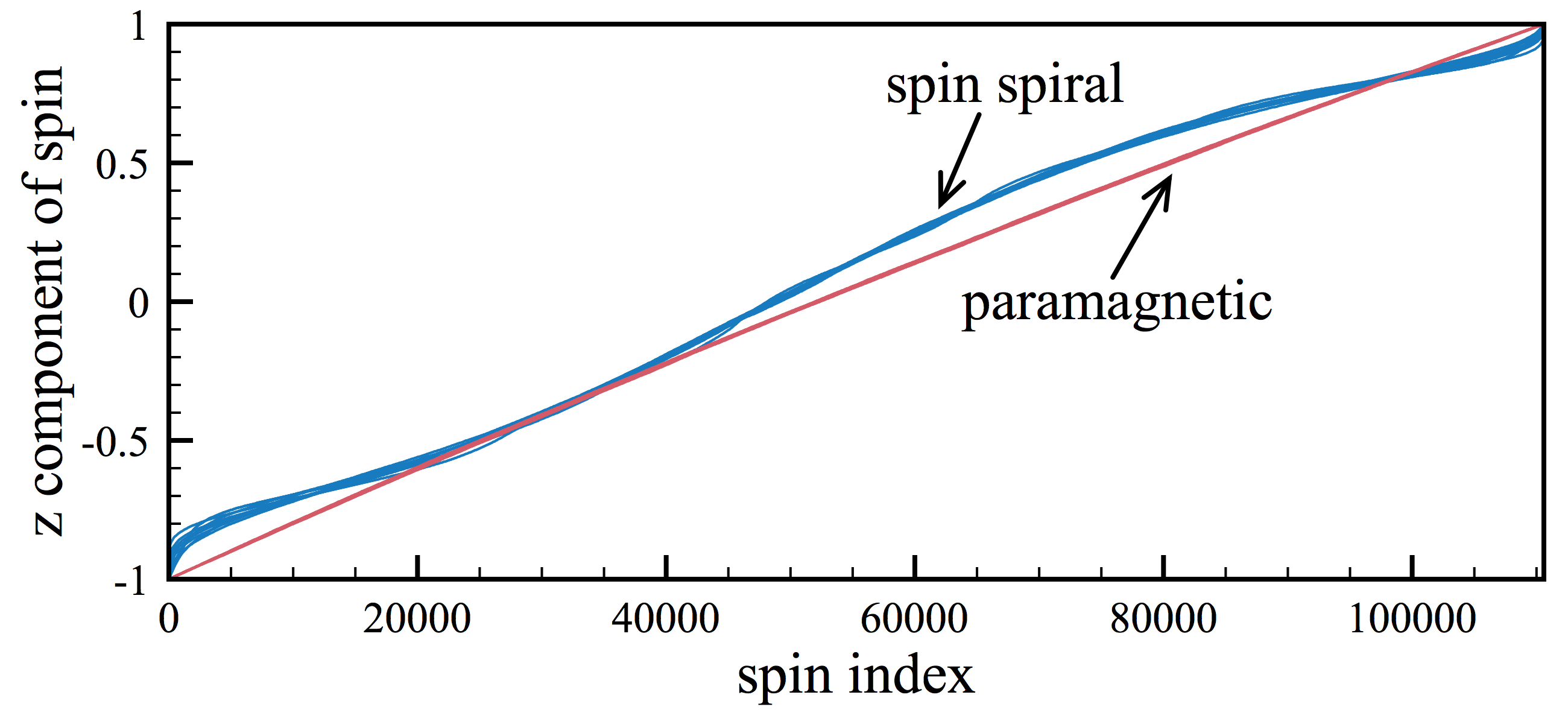}
\caption{Comparison of the magnetization profiles of the spin spiral and paramagnetic configurations obtained with Monte Carlo simulations.}
\label{profile2}
\end{figure}

{\it Profile transformation.}
In contrast to the previous studies \cite{Melko1, Melko2, Wessel,Han1,Sebastian,Hu,Zhang} applying various techniques of the unsupervised learning to spin systems we do not use all the components of spin for profile transformation. It is found that the only information on the $z$ components of the spins is enough for robust classification and recognition of the magnetic structures belonging to different ordered and disordered phases in two- and three-dimensional magnets.  Fig.\ref{profile2} gives comparison of the paramagnetic and spin spiral configurations which have zero total magnetization within the profile approach. The obtained profiles are indeed close to each other, but they still have distinct features. The disordered states are characterized by a pure linear dependence of the magnetization profile. At the same time, spin spiral profile demonstrates oscillations around paramagnetic one, which is related to periodic modulations of the spin structure in the real space.  The fact that the oscillations are non-uniform can be explained as follows.  While the spirals can be formed along different diagonals of the cubic supercell, the magnetization vector for profile transformation is collected from the only $z$ components of spins.

\begin{figure}[b] 
	\center{\includegraphics[width=\columnwidth]{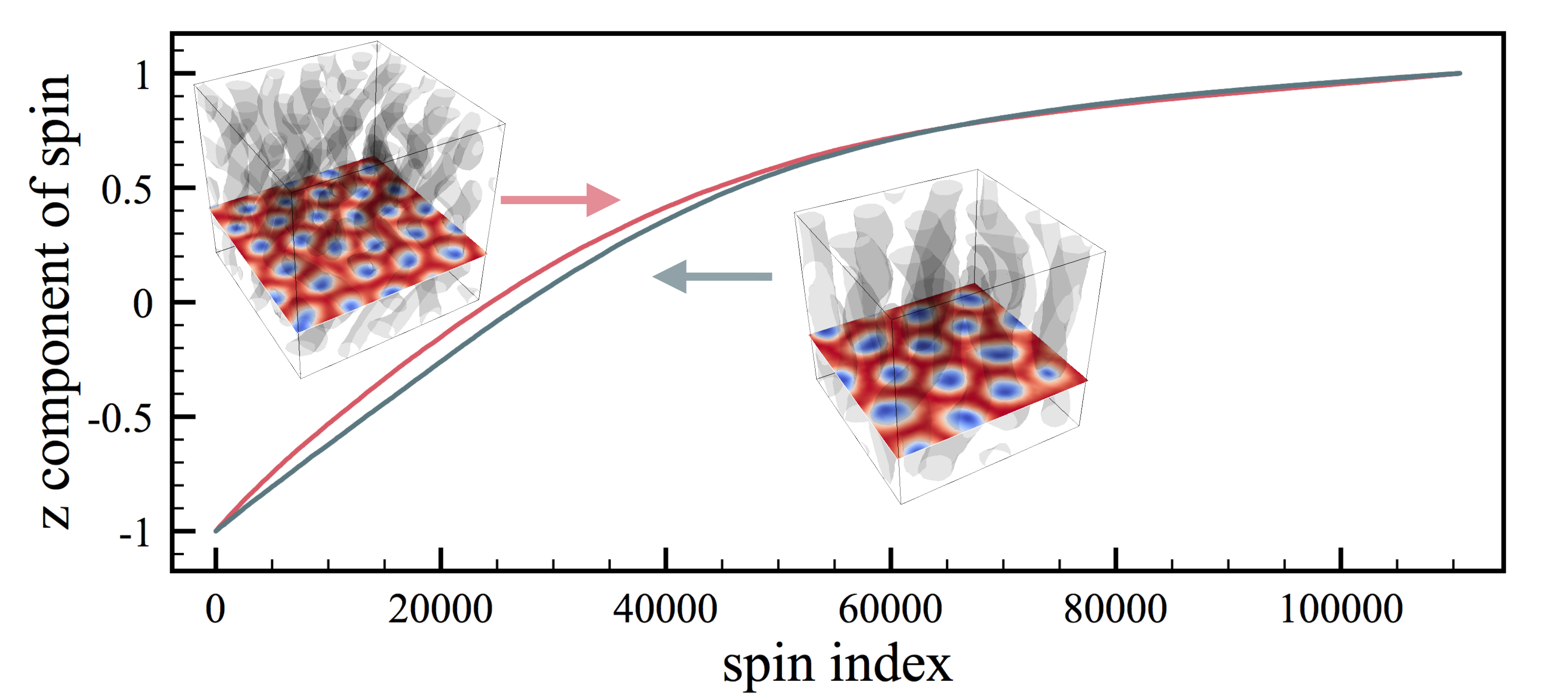} }
    \caption{Skyrmion tubes of different sizes obtained with $|{\bf D}|$ = 1, B = 0.4, T=0.02 (left structure, red profile) and $|{\bf D}|$ = 0.6, B = 0.14, T=0.02 (right structure, gray profile).}
	\label{scale}
\end{figure}

Our approach is numerically and conceptually simpler than the previous ones, since it does not require the calculation of different correlation functions or performing iterative minimization that are needed to be done in standard methods of machine learning. For instance, the principal component analysis (PCA) based on the singular value decomposition projects the high dimensional data (magnetization density) into a low dimensional space of principal components. Then one needs  to perform a dimensional reduction by considering a few principal components \cite{Jarrell}. PCA is used as an initial step in other unsupervised techniques, for instance in t-SNE \cite{Maaten}. At the same time the profile approach requires  sorting of the $z$-component magnetization vector elements in ascending or descending order. We discuss the performance of the t-SNE technique for our problem in the Supplementary Materials \cite{supplementary}. Since we operate only $S^z$ components the profile transformation is computer memory efficient method, which allows to substantially increase the number of spins in a simulated system or set a more fine grid for experimentally measured magnetization.

{\it Scalability.} Importantly, by the construction our approach can capture magnetic structures of different scales but of the same origin. To confirm this we consider two skyrmionic states (Fig.\ref{scale}) stabilized with different Dzyaloshinskii-Moriya interactions. They differ in size and number of the skyrmions. However, the resulting profiles are approximately the same. It ensures the identification of the magnetic objects, regardless of their variations in scale.   

\begin{figure}[b] 
\center 
\includegraphics[width=\columnwidth ,clip]{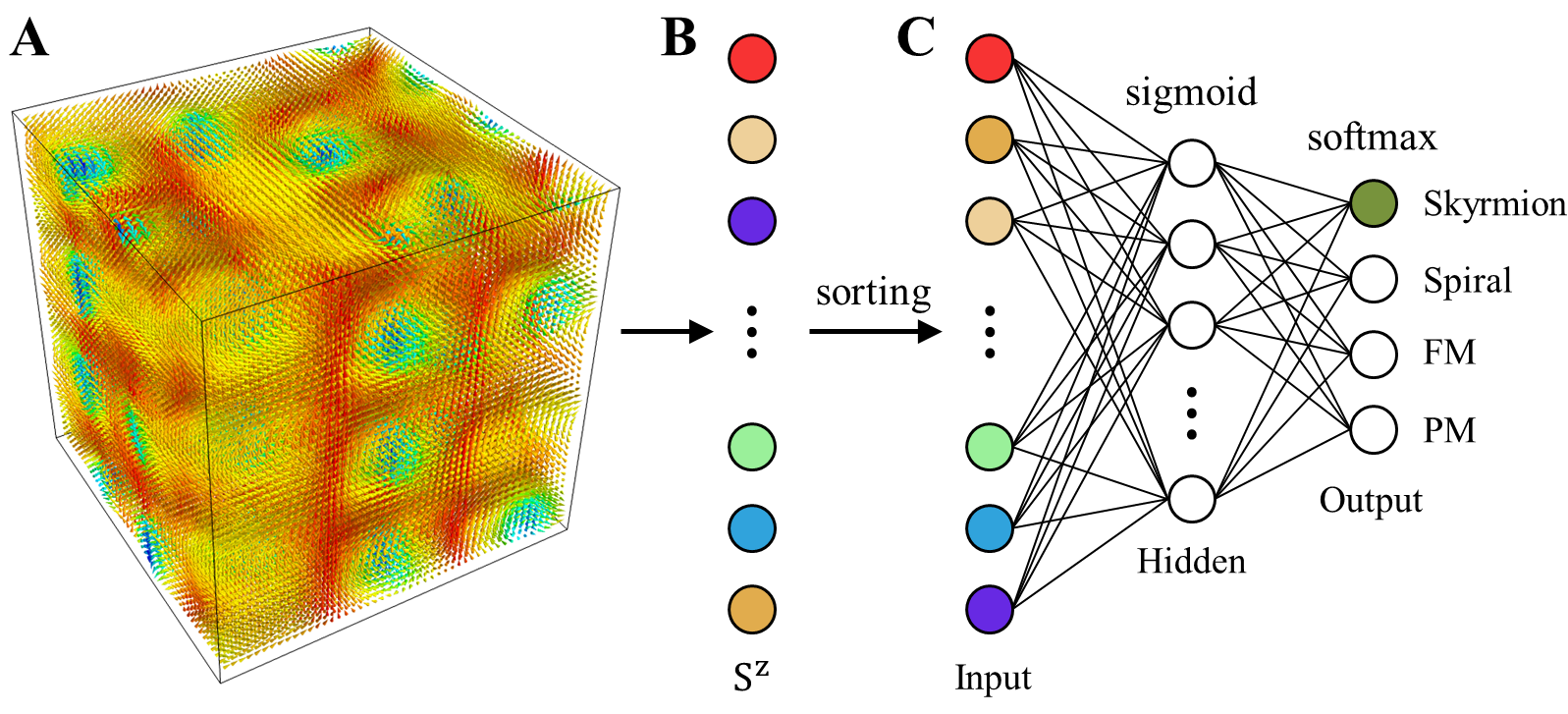}
\caption{Schematic representation of the machine learning process. (A) Example of the skyrmion magnetic structure as obtained from the classical Monte Carlo simulations for a 3D ferromagnet with Dzyaloshinskii-Moriya interaction at finite temperature and magnetic fields. Arrows indicate the spins of individual atoms. (B) Raw vector of $ z $ components of lattice spins. (C) Neural network with single hidden layer of sigmoid neurons we used. The input neurons are equal to the $ z $ components of spins sorted by value.}
\label{neural network}
\end{figure} 

{\it Neural network.} Profile transformation we propose is aimed to distinguish different phases and pattern in unlabelled sets of data on magnetic structures.  Practically, to describe critical areas between different phases, mixed phases and structures one can use a neural network that, as we will show below, can be the simplest one. The feed-forward network scheme we used is presented in Fig.\ref{neural network}. It consists of three layers: input, hidden and output neurons. The values of the input layer neurons are the $S^z$ components of a 3D magnetic configuration. The hidden layer neurons activate by means of sigmoid function. In turn, for output neurons we use the softmax function. The corresponding technical details on machine learning are presented in the Supplementary Materials \cite{supplementary}.

\begin{figure}[b] 
\center 
\includegraphics[width=\columnwidth ,clip]{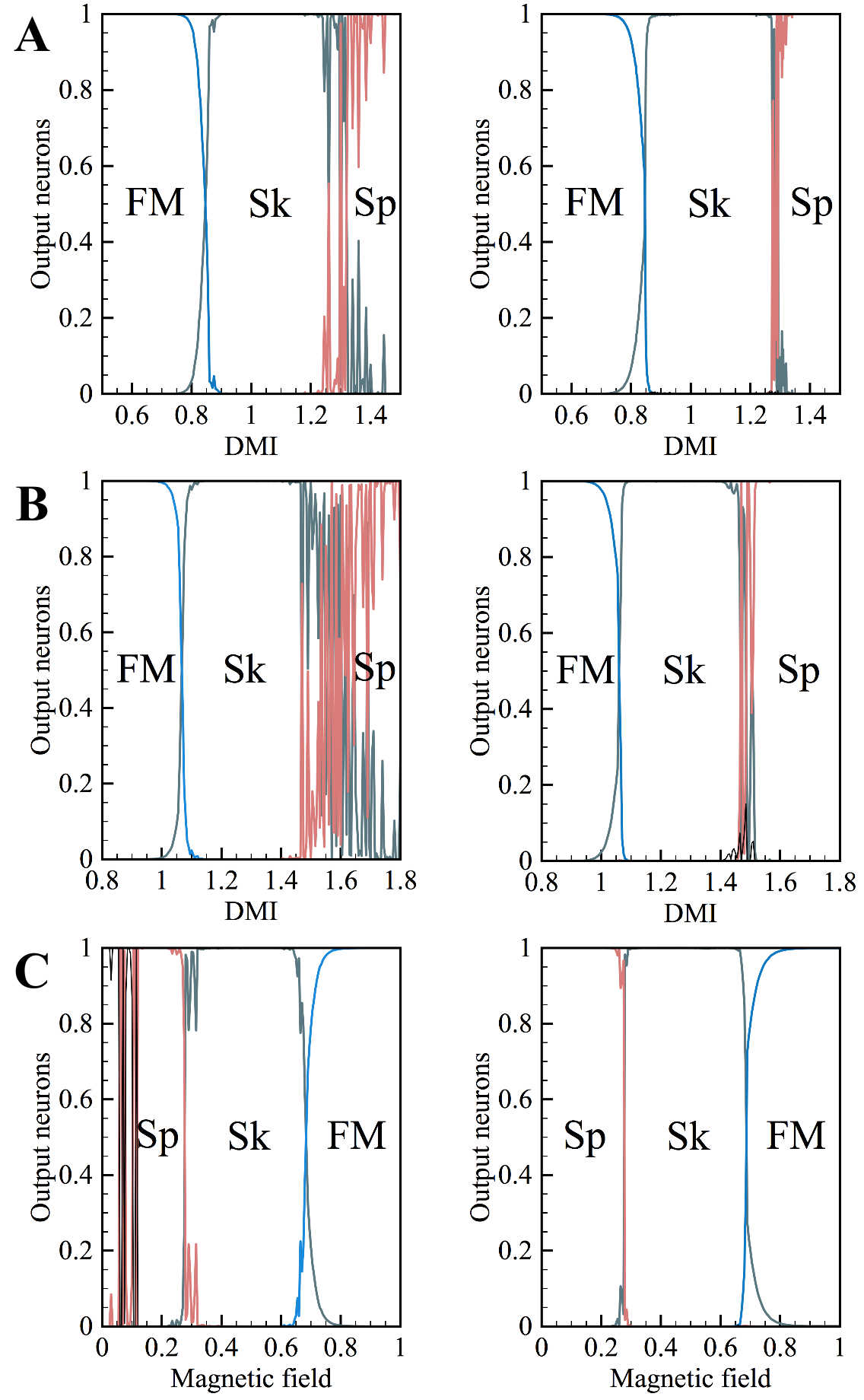}
\caption{Effect of the profile transformation on the results of the supervised learning magnetic phases. FM, Sk and Sp denote ferromagnetic, skyrmion and spin spiral phases, respectively. Left and right panels demonstrate the results obtained without and with preprocessing procedure, sorting of the magnetization vector. Blue, gray, red and black lines correspond to output neuron values for ferromagnetic, skyrmion, spin spiral and paramagnetic phases, respectively. The calculations were performed with (A) $B$ = 0.5 and $T$ = 0.02. (B) $B$ = 0.75 and $T$ = 0.1. (C) $|\bf D|$ = 1 and $T$ = 0.02.  }
\label{outputs}
\end{figure}

For training we used 4600 configurations belonging to pure ferromagnetic, skyrmion,  spin spiral and paramagnetic phases. To define the particular magnetic phase for labelling configurations in the training set we calculated skyrmion numbers and spin structure factors. The detailed information on these calculations is presented in the Supplementary Materials~\cite{supplementary}. To give an accurate description of the magnetic system of 110592 spins we need only 64 neurons in the hidden layer.

Previously, we have shown that the simplest feed-forward network used for classification of the skyrmionic phases in two-dimensional materials learns the magnetization \cite{Iakovlev}. It is due to the construction of the input layer of the network that operates only $z$ components of spins. As a result, the network cannot distinguish the different configurations with similar magnetization, for instance, belonging to spin spiral and paramagnetic phases. If we sort the input vector (profile approach), the network can learn not only magnetization, but also its profile, which solves the problem described above and facilitates the description of the transitional areas between different phases.   

Fig.\ref{outputs} demonstrates the performance of the proposed profile approach. For that we compare the supervised learning results obtained with and without sorting of the input magnetization vector. One can see that the network (left panels) that learns the only total magnetization value of the system without magnetization profile information reveals strong fluctuations of the output values in the critical region. The amplitude of these fluctuations increase at increasing the temperature (Fig.\ref{outputs} B). Such a problem is solved by sorting the input spin values in the magnetization vector on the level of the training and recognition. The phases presented in right panels of Figs. \ref{outputs} A and B  have very clear boundaries. From Fig.\ref{outputs} C one can see that profile transformation (right panel) solves the problem of the misinterpretation of the spin spiral and paramagnetic phases at low magnetic fields. Thus, we provide a more accurate description of the transitional areas than those obtained with convolutional neural networks in Ref.\cite{Han1} for two-dimensional magnets and characterized by strong fluctuations. 

 \begin{figure}[t] 
\center 
\includegraphics[width=\linewidth ,clip]{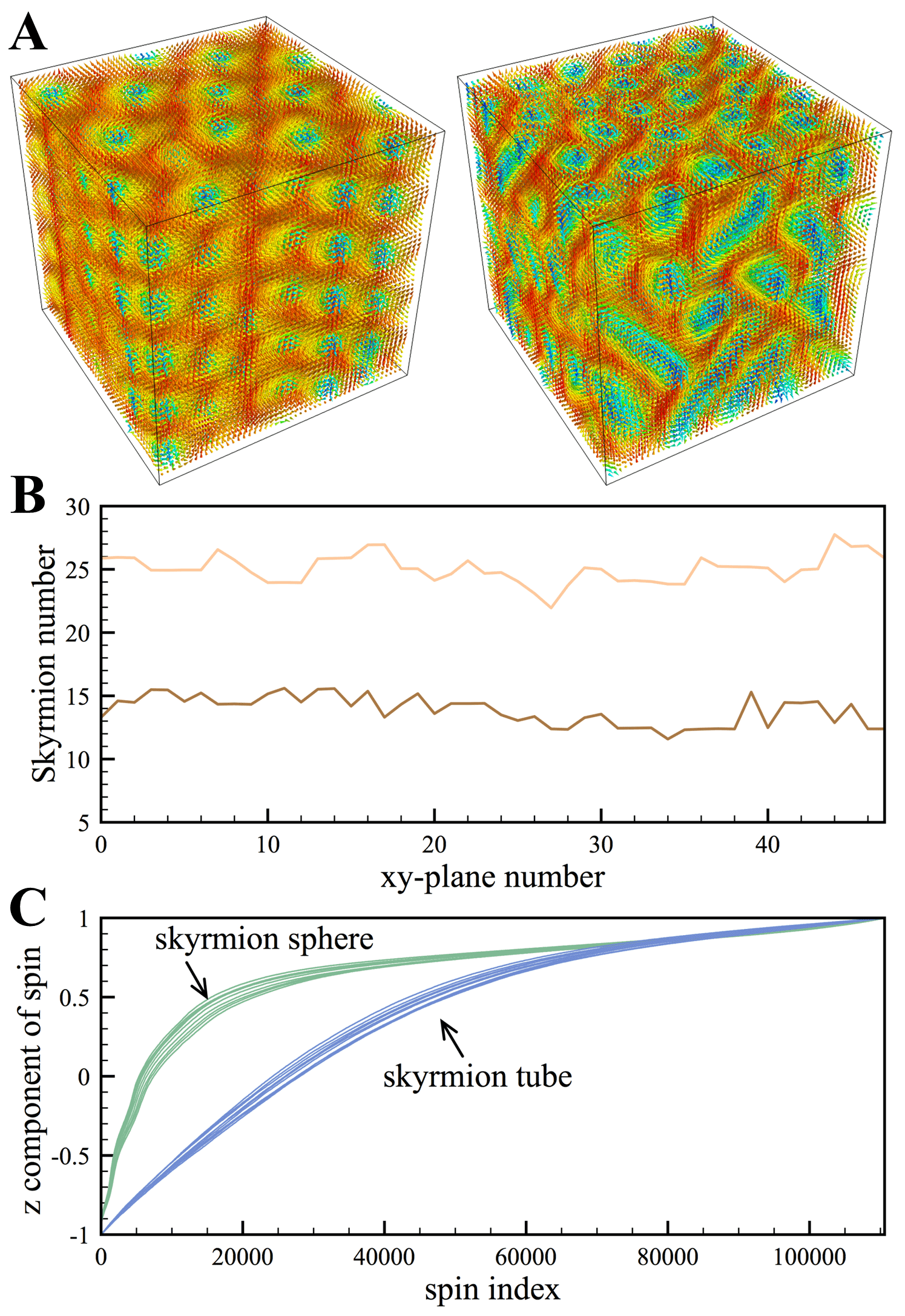}
\caption{(A) Examples of the skyrmion tube and sphere lattices obtained with $B$ = 0.6, $|{\bf D}|$ = 1, $T$ = 0.02 (spheres) and $B$ = 0.4, $|\bf D|$ = 1, $T$ = 0.02 (tubes).  (B) The corresponding skyrmion numbers with the resolution on the $xy$ planes along $z$ direction. (C) Comparison of the profiles of skyrmion tube and sphere structures sets.}
\label{Q_z}
\end{figure}

{\it Detection of the skyrmionic patterns.} Having classified magnetic phases by means of the profile transformation we are going to demonstrate that our approach can unveil different patterns in data on spin structures belonging to the same phase. Previous theoretical studies \cite{3dskyrmion1, 3dskyrmion2, 3dskyrmion3, Han} on topologically-protected structures have revealed the formation of different skyrmion states in 3D magnets. They are spherical skyrmions and tubular skyrmions formed at high and low magnetic fields, respectively. Alternatively, one can change the Dzyaloshinskii-Moriya interaction on the level of the spin Hamiltonian to switch between skyrmionic structures of different types. The examples are presented in Fig.\ref{Q_z}. The calculated skyrmion numbers as a function of the layer number $z$ shows an oscillating behaviour in all the cases and thus can not be used as indicator for particular skyrmionic state. 

Our approach provides an elegant solution of this problem, the profiles corresponding to skyrmion tubes and spheres are different. In the case of the skyrmion sphere-like structures we observe plateaus in the profile for the negative $z$ components of spins. These plateaus are related to the fact that skyrmionic spheres have almost the same spin texture with similar values of $S^z$ on each particular distance from the centre. The mixed sphere-tube skyrmionic configurations are in the gap between these two extreems and to recognize them it is better to use neural network approach as it was done in the case of the phase classification.

{\it Conclusion.}
We show that operating $z$ components of spins of a three-dimensional magnet one can distinguish different magnetic phases formed at finite temperatures and magnetic fields. Importantly, the same approach can be applied to recognize various patterns in magnetic structure data corresponding to the same phase.
In comparison with other methods of unsupervised learning (PCA, t-SNE) assuming a high level of abstraction, in the profile approach the principal quantity is magnetization that has a direct physical meaning. 

{\it Acknowledgments.}
This work was supported by the Russian Science Foundation Grant 18-12-00185.

\newpage

\widetext
\begin{center}
\textbf{\large Supplemental Material: Profile approach for recognition of three-dimensional magnetic structures}
\end{center}
\setcounter{equation}{0}
\setcounter{figure}{0}
\setcounter{table}{0}
\setcounter{page}{1}
\makeatletter
\renewcommand{\theequation}{S\arabic{equation}}
\renewcommand{\thefigure}{S\arabic{figure}}
\renewcommand{\bibnumfmt}[1]{[S#1]}
\renewcommand{\citenumfont}[1]{S#1}

\section{Model and Methods}
The spin Hamiltonian, Eq.(1) in the main text was solved by using the classical Monte Carlo approach. The spin update scheme is based on the Metropolis algorithm. The systems in question are gradually (200 temperature steps) cooled down from high temperatures (${\rm T}\sim 3J$) to the required temperature. Each temperature step run consists of $1.5\times10^{6}$ Monte Carlo steps.

To define the particular magnetic phase for labelling configurations in the training set we calculated some observables. They are spin structural factors and skyrmion numbers. The expressions for spin structure factors are given by
\begin{equation}\label{hi1}
\chi_\parallel(\textbf{q})=\frac{1}{N}\left\langle\left|\sum_{i}S^z_ie^{-i\textbf{q}\textbf{r}_i}\right|^2\right\rangle,
\end{equation}

\begin{equation}\label{hi2}
\chi_\perp(\textbf{q})=\frac{1}{N}\left\langle\left|\sum_{i}S^x_ie^{-i\textbf{q}\textbf{r}_i}\right|^2+\left|\sum_{i}S^y_ie^{-i\textbf{q}\textbf{r}_i}\right|^2\right\rangle,
\end{equation}
where $\textbf{q}$ is the reciprocal space vector, $S_i^\alpha$ ($\alpha = \{x,y,z\}$) is the projection of the $i$th spin and $\textbf{r}_i$ is the radius vector for the $i$th site.

\begin{figure}[b]
\begin{minipage}[h]{\qcol}
\center{\includegraphics[width=1\columnwidth]{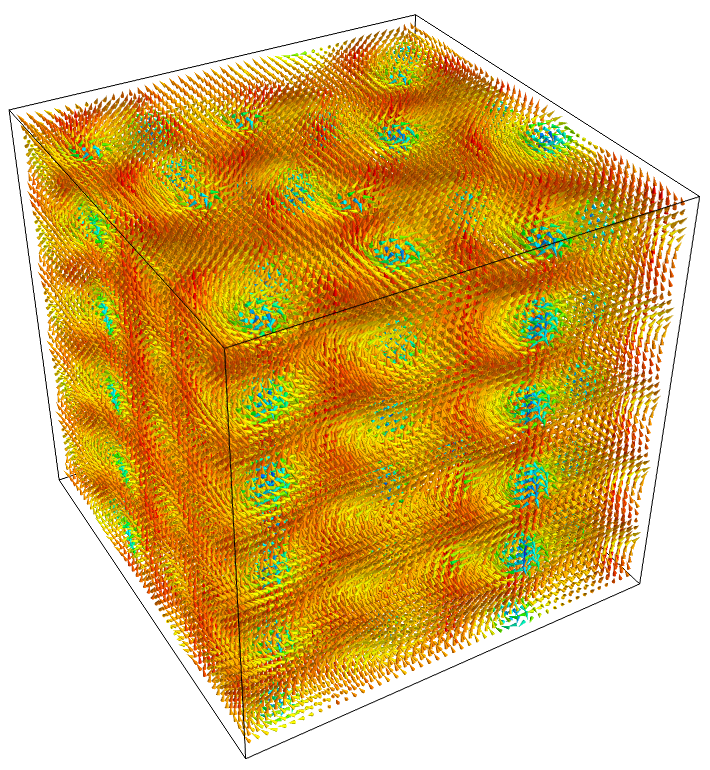}}
\end{minipage}
\begin{minipage}[h]{\qcol}
\center{\includegraphics[width=1\columnwidth]{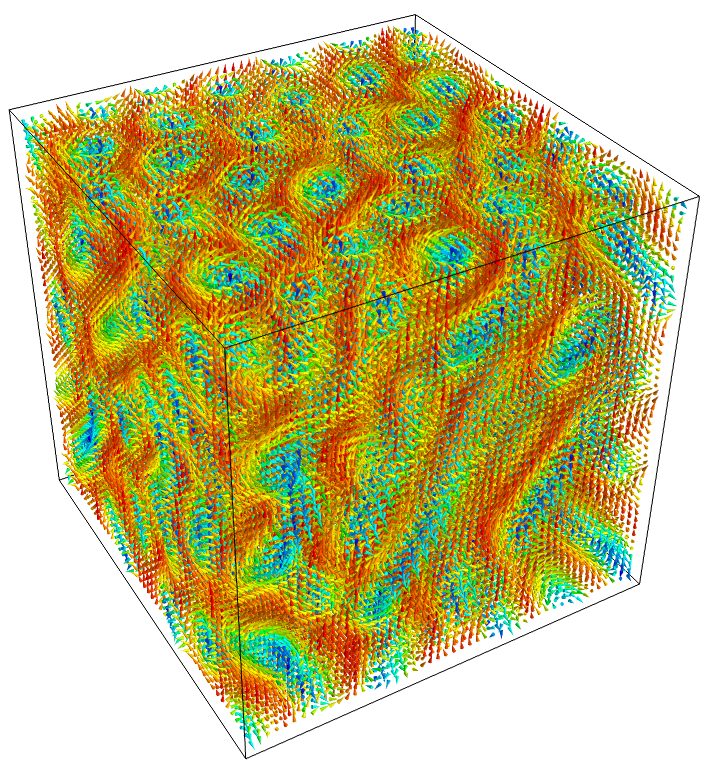}}
\end{minipage}
\begin{minipage}[h]{\qcol}
\center{\includegraphics[width=1\columnwidth]{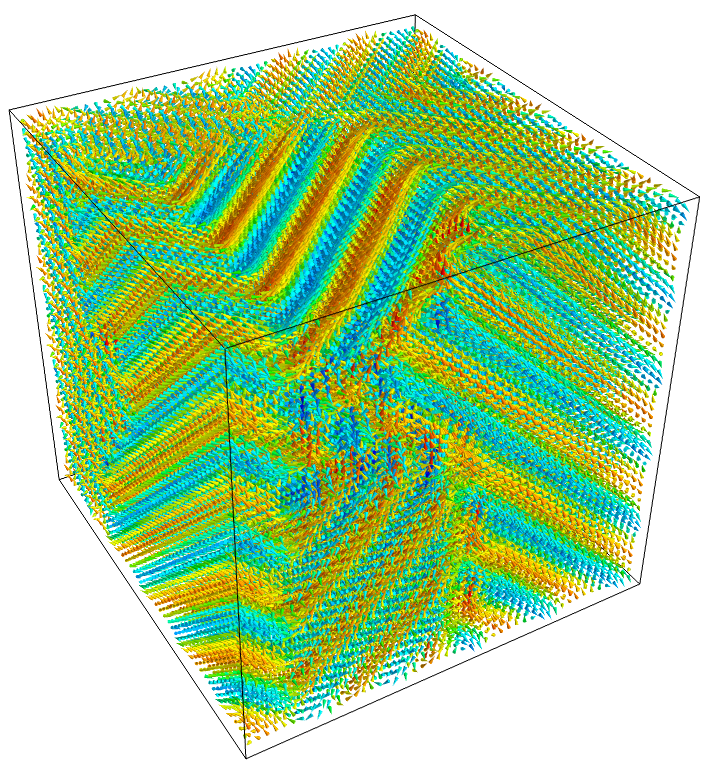}}
\end{minipage}
\begin{minipage}[h]{\qcol}
\center{\includegraphics[width=1\columnwidth]{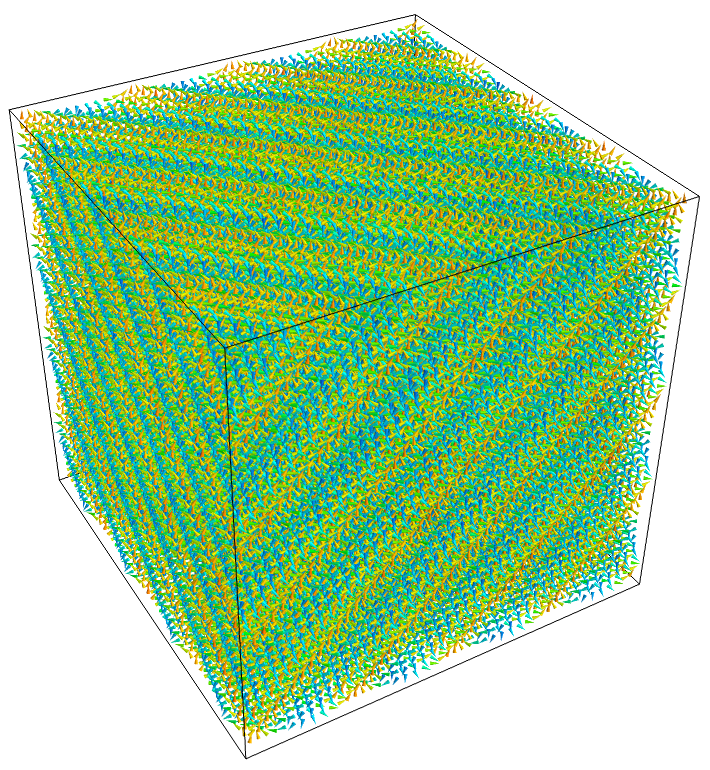}}
\end{minipage}
\begin{minipage}[h]{\qcol}
	\center{\includegraphics[width=1\columnwidth]{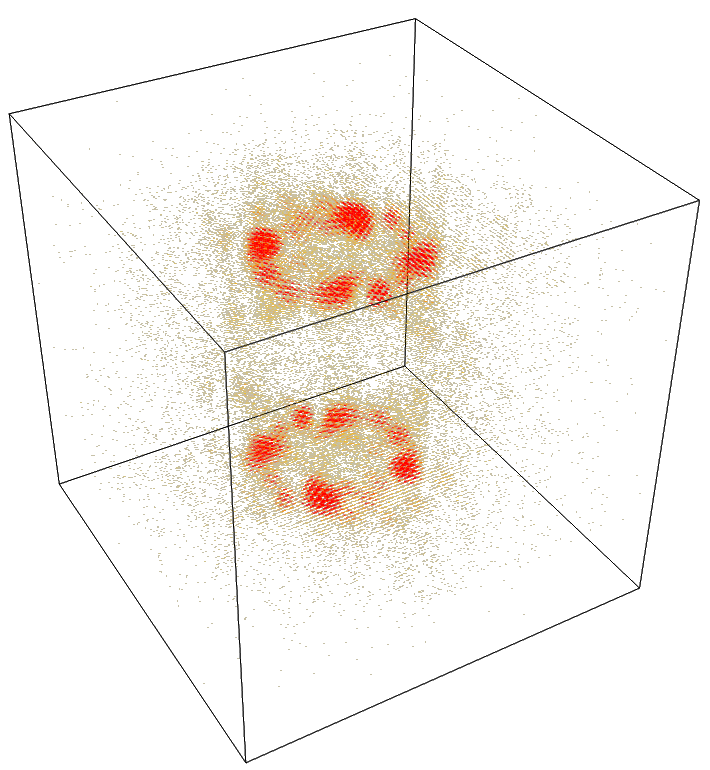}}
\end{minipage}
\begin{minipage}[h]{\qcol}
	\center{\includegraphics[width=1\columnwidth]{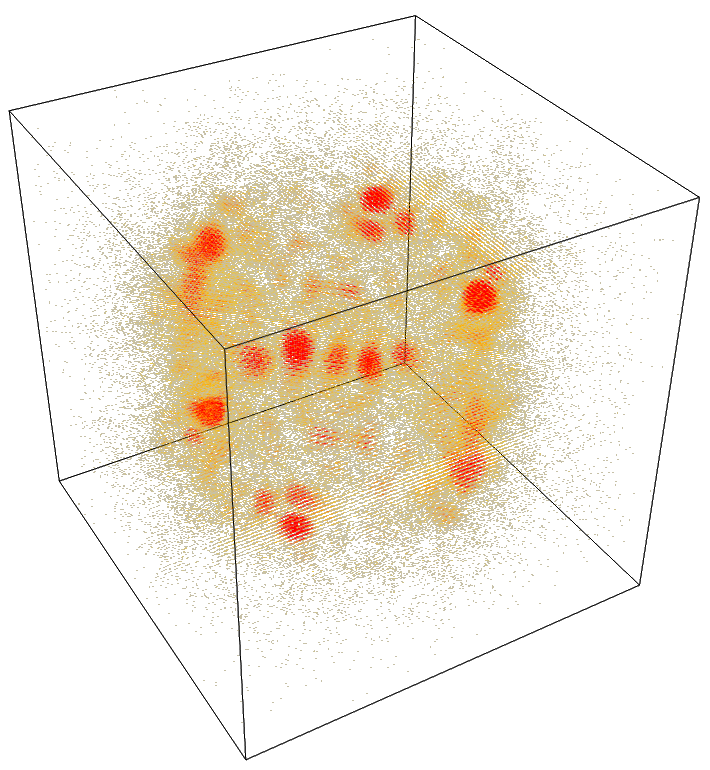}}
\end{minipage}
\begin{minipage}[h]{\qcol}
	\center{\includegraphics[width=1\columnwidth]{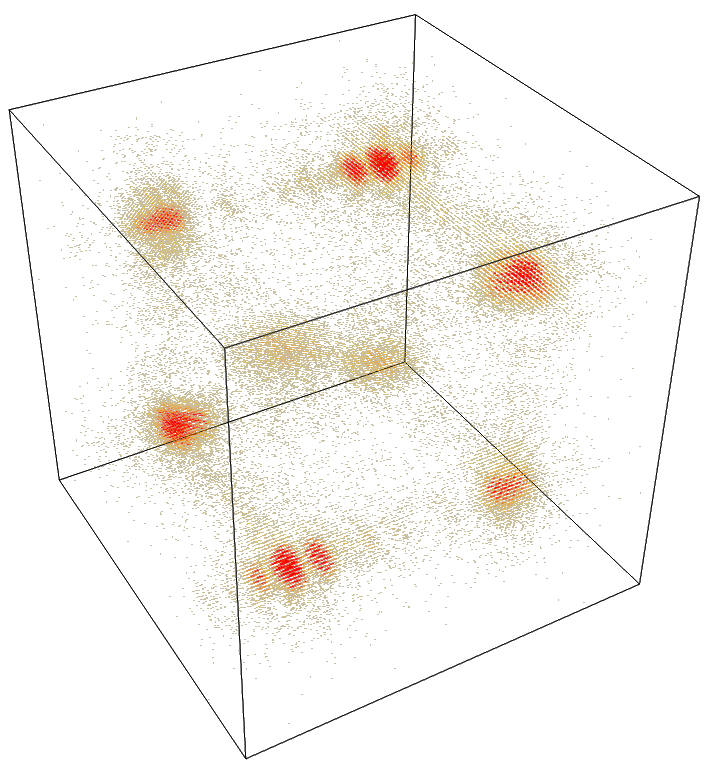}}
\end{minipage}
\begin{minipage}[h]{\qcol}
\center{\includegraphics[width=1\columnwidth]{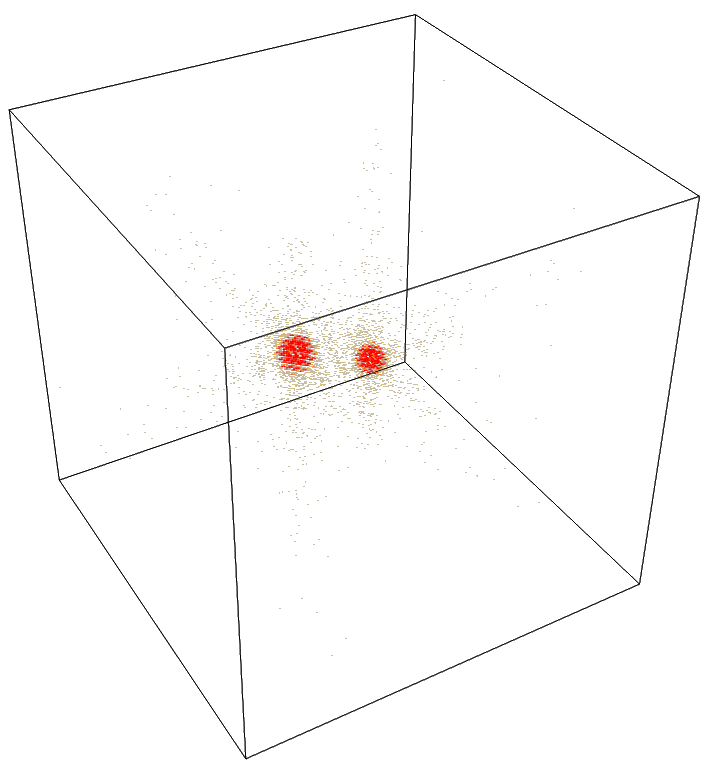}}
\end{minipage}
\caption{Simulated skyrmion (A and B) and spin spiral (C and D) configurations (top panels) and the corresponding spin structure factors (bottom panels). The configurations were obtained with the following parameters: $|\bf D|$ = 0.93, $B$ = 0.56 (A), $|\bf D|$ = 1.15, $B$ = 0.55 (B), $|\bf D|$ = 1.35, $B$ = 0.2 (C), $|\bf D|$ = 1, $B$ = 0.01 (D). All configurations were obtained at $T$ = 0.02. Corresponding values are given in units of $J$.}
\label{struct}
\end{figure}

Following Ref.~\onlinecite{Han_s} we define the topological charge for each $xy$-plane along $z$ direction
\begin{equation}\label{Q}
Q(z) =\frac{1}{4\pi}\sum_{l}A_{l},
\end{equation}
where $A_l$ is the solid angle subtended by three spins located at the vertices of an elementary triangle $l$ \cite{Berg_s},
\begin{equation}\label{A_l}
A_l=2\arccos\left(\frac{1+\textbf{S}_i\!\cdot\textbf{S}_j+\textbf{S}_j\!\cdot\textbf{S}_k+\textbf{S}_k\!\cdot\textbf{S}_i}{\sqrt[]{2(1+\textbf{S}_i\!\cdot\textbf{S}_j)(1+\textbf{S}_j\!\cdot\textbf{S}_k)(1+\textbf{S}_k\!\cdot\textbf{S}_i)}}\right).
\end{equation}
The sign of $A_l$ in Eq.~\eqref{Q} is determined as $sign(A_l)=sign(\textbf{S}_i\cdot[\textbf{S}_j\times\textbf{S}_k])$. Importantly, we do not consider the exceptional configurations for which
\begin{align}
&{\bf S}_{i} \cdot [{\bf S}_{j} \times {\bf S}_{k}] = 0, \\
&1+ {\bf S}_{i} \!\cdot {\bf S}_{j} + {\bf S}_{j} \!\cdot {\bf S}_{k} + {\bf S}_{k} \!\cdot {\bf S}_{i} \le 0. \notag
\end{align} 

By the example of the configurations plotted in Fig.\ref{struct} we demonstrate the complexity of the problem of the three-dimensional magnetic structure classification. One can see that existence of broken spiral structures leads to multiple maximum intensities in the spin structure factors, which complicates the classification. 

\section{Neural network details}
\begin{figure}[t] 
	\center 
	\includegraphics[width=0.45\columnwidth]{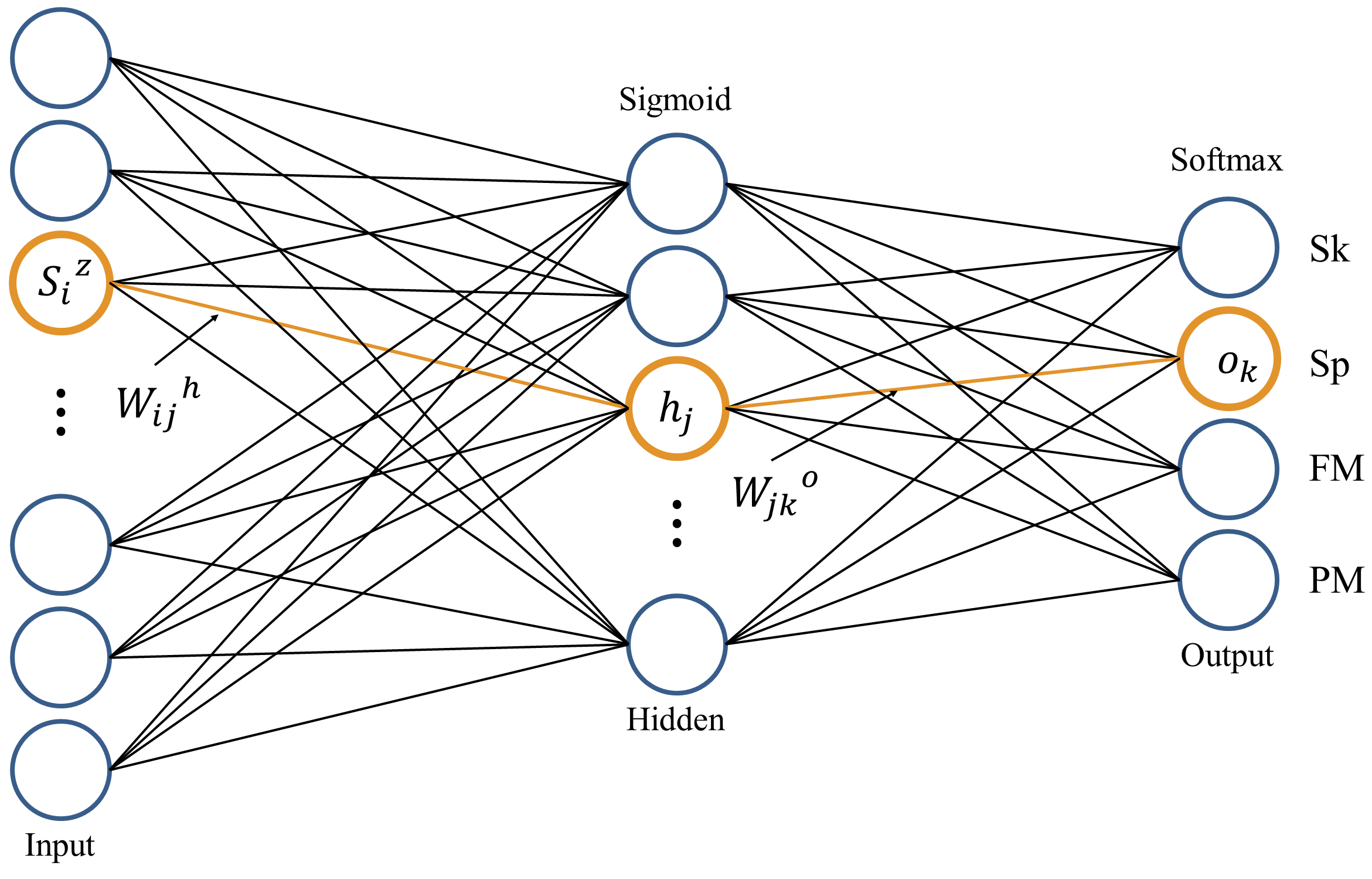} 
	\caption{Schematic representation of the neural network with single hidden layer. We used sigmoid as an activation function of the hidden neurons and softmax for the output ones. All the notations are described in the text.}
	\label{sup_netw} 
\end{figure}

As an input for our feed-forward network (Fig.\ref{sup_netw}) we used the $S^z$ components of a three-dimensional magnetic configuration. The hidden layer neurons activate by means of sigmoid function,
\begin{eqnarray}
h_j = sigmoid (x_j) = \frac{1}{1+e^{-x_j}},
\end{eqnarray}
\begin{equation}\label{hi}
x_j= \frac{1}{N} \sum_{i=1}^{N}S_i^z W^h_{ij},
\end{equation}
where $S_i^z$ is the value of $i$th input neuron, $W^h_{ij}$ --- weight between the $i$th input neuron and $j$th hidden neuron, $N=L\times L\times L$ --- number of the input neurons. The normalization factor in the second equation is required in order to shift the input value into the range where $sigmoid(x_j)\in[0; 1]$. It is very important especially at the beginning of the learning process when we randomly initialize all weights in range $[-1; 1]$. Without normalization we would obtain  $h_j$ that are equal to 1 or 0 because of the huge dimensionality of the input vector. It leads to the situation when weights between the hidden and output neurons become the only parameters that affect the result.

In turn, for output neurons we use the softmax function that is given by 
\begin{eqnarray}
o_k = softmax(y_k) = \frac{e^{y_k}}{\sum_{n = 1}^{N_o}e^{y_n}},
\end{eqnarray}
\begin{equation}
y_k= \sum_{j=1}^{N_h}h_j W^o_{jk},
\end{equation}
where $N_o$ is the number of the output neurons, $N_h$ --- number of the hidden neurons, $W^o_{jk}$ --- weight between the $j$th hidden neuron and $k$th output neuron.

During the learning process, we randomly chose 10\% of training set for cross-validation to avoid overfitting and define the stopping point where error is less than the required value. To evaluate the error we used cross entropy loss function is given by
\begin{equation}\label{Error}
L(o^{ideal},o^{actual})= -\sum_{k=1}^{N_o}o_k^{ideal}\log o_k^{actual},
\end{equation}
where $o^{ideal}$ represents the training labels and $o^{actual}$ is the calculated values of the output neurons.

The network optimization was fulfilled through back-propagation method \cite{backprop_s} by means of the stochastic gradient descent with momentum. We used the following expressions for new weights in order to avoid getting stuck in a local minima 
\begin{equation}\label{wi}
W^{(l)}= W^{(l-1)}+\Delta W^{(l)},
\end{equation}
\begin{equation}\label{dwi}
\Delta W_{jk}^{o(l)}=  \alpha\delta o_k h^{out}_j+\mu\Delta W_{jk}^{o(l-1)},
\end{equation}
\begin{equation}\label{dwi}
\Delta W_{ij}^{h(l)}=  \alpha\delta h^{out}_j S^z_i+\mu\Delta W_{ij}^{h(l-1)},
\end{equation}
where $\mu$ is the momentum and $\alpha$ is the learning rate. These parameters can be chosen by trial and error (in our work we used $\mu=0.3$, $\alpha=0.8$). $\delta o_k$ and $\delta h^{out}_j$ are given by 
 
\begin{equation}\label{do}
\delta o_k= (o_k^{ideal}-o_k),
\end{equation}
\begin{equation}\label{dh}
\delta h^{out}_j= h^{out}_j(1-h^{out}_j)\sum_{k=1}^{N_o}W^o_{jk}\delta o_k.
\end{equation}

\begin{figure}[b] 
	\includegraphics[width=0.5\columnwidth ,clip]{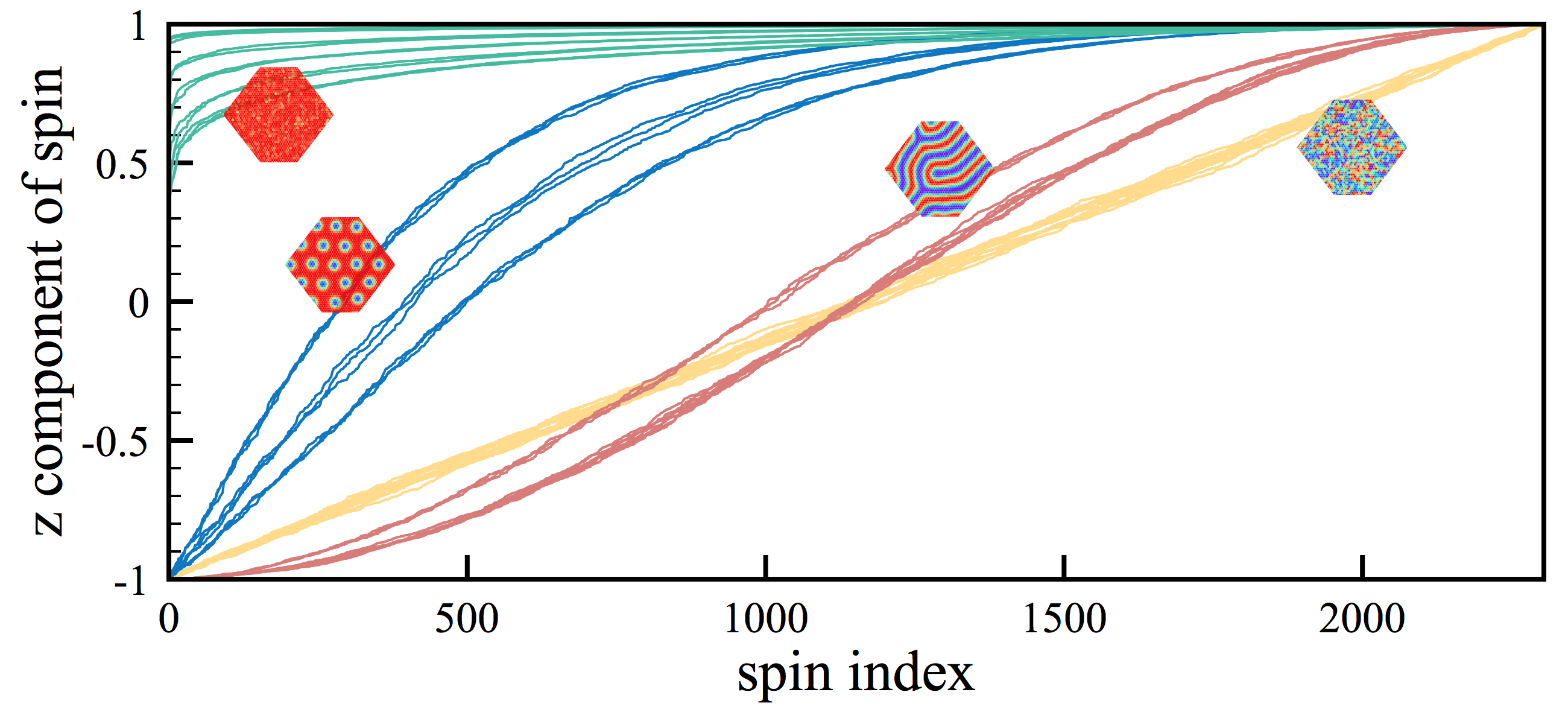}
	\caption{Magnetization profiles of configurations belonging to paramagnetic (yellow lines), spin spiral (red lines), skyrmion (blue lines) and ferromagnetic (green lines) phases obtained with the spin Hamiltonian on the two-dimensional triangular lattice.}
	\label{profile_triangular}
\end{figure}

\section{Two-dimensional lattices}

In order to verify the universality of the proposed profile approach with respect to the dimensionality of the system in question and lattice structure we have analyzed the Monte Carlo results obtained for spin Hamiltonian on the two-dimensional triangular lattice. They are presented in Fig.~\ref{profile_triangular}. Similar to the results presented in the main text there are clearly seen patterns in magnetization profiles, which facilitates the classification of the magnetic structure. Importantly, the behaviour of the magnetic profiles is almost the same as in the case of the three-dimensional magnets with cubic lattice (Fig.1, main text). It means that profile is an universal property of a particular phase that does not depend on the dimensionality and lattice structure.     

\section{t-SNE results}
Previous works \cite{Melko1_s,Melko2_s, Hu_s, Wessel_s,Sebastian_s,Evert_s} on the classification of the magnetic phases in two-dimensional magnets have demonstrated that unsupervised machine learning technique such as PCA, t-SNE and autoencoder can capture phase transitions and critical points of the two-dimensional Ising  and XY models as well their different extensions.

\begin{figure}[!h] 
  \includegraphics[width=0.75\columnwidth ,clip]{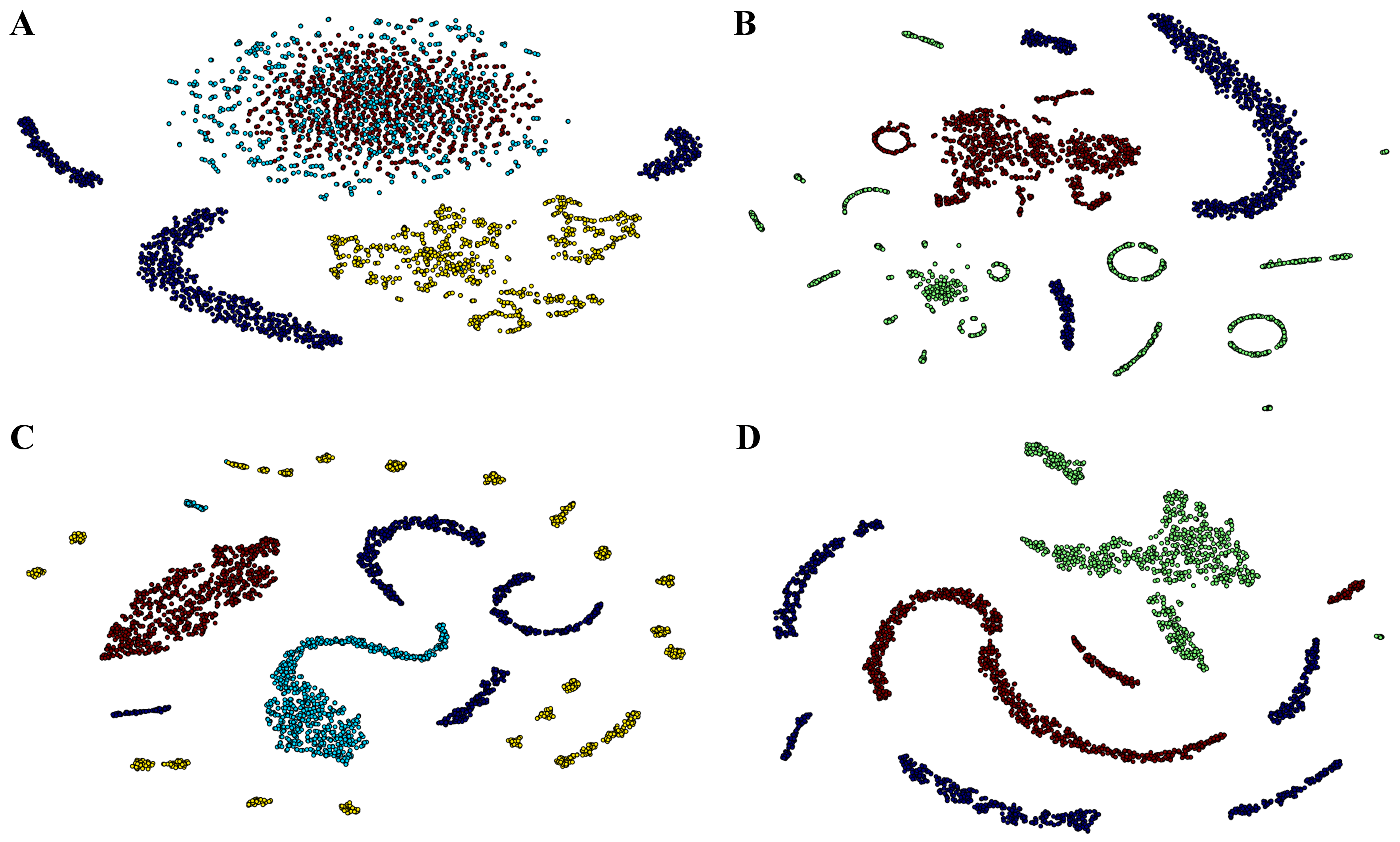}
	\caption{(Left) Two-dimensional t-SNE representation of the 3524 Monte Carlo configurations on the triangular $L = 48$ lattice. (Right) Two-dimensional t-SNE representation of the 3000 Monte Carlo configurations on the square $L = 48$ lattice. (A and B) raw input vector, (C and D) sorted input vector. Color labels different types of magnetic configurations: skyrmions, ferromagnetic, spin spirals and paramagnetic.}
	\label{tsne}
\end{figure}

To demonstrate the complexity of the classification problem in the case of the spin Hamiltonian containing Dzyaloshinskii-Moriya interaction favoring stabilization of topological multispiral states we use t-SNE technique to visualize magnetic configurations which belong to pure ferromagnetic, skyrmion, paramagnetic and spin spiral phases. Here we concentrate on the two-dimensional square and triangular lattices \cite{Iakovlev_s}. From Fig.\ref{tsne} one can see that there is a large overlap between two phases in the case of the triangular lattice (Fig.\ref{tsne} A) and separation of the manifolds of the configurations belonging to the same phase of the spin Hamiltonian on the square lattice denoted with green color (Fig.\ref{tsne} B). 

The situation can be improved by sorting the magnetization vectors as a preconditioning procedure for t-SNE technique (Fig.\ref{tsne} C and D). In this case  there is a clear separation of different classes.  In all the cases the number of the classes in the low-dimensional representation of the data is much larger than number of magnetic phases. 
It means that t-SNE can not be used to generate the labels for supervised learning in our case.    

\end{document}